\documentclass[journal]{IEEEtran}
\usepackage{amsmath,amsfonts}
\usepackage{algorithm}
\usepackage{array}
\usepackage[caption=false,font=normalsize,labelfont=sf,textfont=sf]{subfig}
\usepackage{textcomp}
\usepackage{stfloats}
\usepackage{url}
\usepackage{verbatim}
\usepackage{graphicx}
\usepackage{cite}
\usepackage{tikz}
\usepackage[thinc]{esdiff}
\usepackage{amssymb}
\usepackage{algpseudocode}

\begin{document}

\title{Gamma-Based Statistical Modeling for Extended Target Detection in mmWave Automotive Radar}

\author {Vinay Kulkarni, V. V. Reddy\\
\footnotesize \textit {\scriptsize {vinay.kulkarni@iiitb.ac.in, vinod.reddy@iiitb.ac.in}}\\
\footnotesize \textit{ International Institute of Information Technology,
Bengaluru, India} \\
}

\maketitle

\begin{abstract}
Millimeter-wave (mmWave) radar systems, owing to their large bandwidth, provide fine range resolution that enables the observation of multiple scatterers originating from a single automotive target—commonly referred to as an extended target. Conventional CFAR-based detection algorithms typically treat these scatterers as independent detections, thereby discarding the spatial scattering structure intrinsic to the target. To preserve this scattering spread, this paper proposes a Range-Doppler (RD) segment framework designed to encapsulate the typical scattering profile of an automobile. The statistical characterization of the segment is performed using Maximum Likelihood Estimation (MLE) and posterior density modeling based on the Gamma distribution, facilitated through Gibbs Markov Chain Monte Carlo (MCMC) sampling.A skewness-based test statistic, derived from the estimated statistical model, is introduced for binary hypothesis classification of extended targets. Additionally, the paper presents a detection pipeline that incorporates Intersection over Union (IoU) and segment centering based on peak response, optimized to work within a single dwell. Extensive evaluations using both simulated and real-world datasets demonstrate the effectiveness of the proposed approach, underscoring its suitability for automotive radar applications through improved detection accuracy.
\end{abstract}


\section{Introduction}

\IEEEPARstart Automotive radars are increasingly operating at mmWave and microwave frequencies, resulting in a substantial reduction in the operating wavelength ($\lambda$), often much smaller than typical object dimensions. As a result, automotive targets are typically categorized within the optical scattering region~\cite{skolnik1980introduction}, where object sizes exceed 10–100 times the wavelength. Targets in this regime, referred to as large or extended targets, exhibit radar reflections dominated by geometric and specular scattering mechanisms, highly dependent on the shape, surface characteristics, and material composition of objects such as cars, SUVs, and trucks. Although the shorter wavelengths enhance resolution and offer greater bandwidth, they also cause the target size to exceed the radar resolution cell, leading to complex scattering behavior and posing significant challenges for reliable target detection~\cite{wang2016spatial,barton1997radar}.

In practical automotive applications, radar sensors are strategically positioned to maximize coverage. Long-range and mid-range radars are mounted at the front and rear of the ego vehicle to detect objects at greater distances, while short-range radars are placed at the corners to assist with parking and pedestrian detection. These radar systems rely on Range-Doppler (RD) and/or Range-Angle (RA) maps for target detection, employing binary statistical hypothesis testing, where target absence and presence correspond to the Null and Alternate hypothesis, respectively ~\cite{kay1993fundamentals}. To ensure reliable detection, radar signal processing (RSP) employs constant false-alarm rate (CFAR) detectors, which dynamically adjust thresholds to maintain a consistent false alarm rate ~\cite{richards2014fundamentals, richards2010principles, skolnik1980introduction}. Among various CFAR techniques, ordered-statistics (OS)-CFAR is widely used in automotive radar due to its robustness in varying noise environments~\cite{sun2020mimo}. Given the need for accurate detection~\cite{karangwa2023vehicle} with minimal dwell time, low latency, and self-contained efficiency, reducing reliance on subsequent processing modules such as association and tracking~\cite{pearce2023multi} has become a key priority for modern automotive radar systems.

In the optical scattering region, multiple radar reflections cause significant scattering spread across both RD and RA maps. In automotive scenarios, a car may occupy several cells in both range and Doppler, forming a large target spread in the RD domain. A similar spread is also observed when two or more closely-situated targets exist in RD map. This challenges CFAR detectors, which operate on a cell-by-cell basis, leading to an increased number of detections. Consequently, a single object may appear as multiple targets or produce redundant detections, complicating the association and tracking processes.

Statistical hypothesis testing methods, such as the Generalized Likelihood Ratio Test (GLRT), employ the maximum likelihood estimate of unknown parameters under both the null and alternative hypothesis at the target range cell for detection~\cite{aubry2015radar}. The adaptive two-step GLRT first estimates the sample covariance matrix (SCM) using secondary data before maximizing the likelihood ratio~\cite{aubry2014radar,fuhrmann1992cfar}. Variants like the Rao and Wald tests rely on the Fisher Information Matrix (FIM)\cite{liu2014fisher} for symmetric random parameters\cite{liu2022multichannel}. These methods typically assume a known noise covariance. At the same time, adaptive approaches~\cite{brennan1973theory,kelly1960detection} depend entirely on secondary data, making them susceptible to performance degradation when secondary data is limited, as seen in automotive scenarios with significant scattering spread.

CFAR-based adaptive detection techniques, such as truncated statistics CFAR (TS-CFAR)\cite{tao2015robust}, mitigate the impact of statistical outliers in reference windows. Variance-based approaches, including quantile truncated statistics (QTS) and QTS with maximum likelihood estimation (QTS-MLE), further enhance robustness\cite{zhou2022robust}. Comprehensive CFAR (Comp-CFAR) leverages the central limit theorem and log compression for non-coherent accumulation while incorporating a protection window for cell-averaging CFAR, improving adaptability to normally distributed target data~\cite{liu2019research}. Similarly, Greatest of Secondary Detection (GOSD)-CFAR addresses target masking in mmWave radar, preserving detection probability ($P_{\mathrm{D}}$) without increasing the probability of false alarm ($P_{\mathrm{FA}}$)~\cite{qin2020novel}. However, challenges persist in optimizing convolution truncation, refining reference window selection, and mitigating compression-induced loss of target scattering spread in automotive radar. Area-Based Combination (ABC) and Area-Based Distribution (ABD) CFAR refine detection by applying fixed-kernel convolutions over reference windows instead of individual reference cells to determine detection thresholds~\cite{wei2022area}. While these methods improve spatial target characterization, selecting the optimal kernel size remains a critical challenge.

Stochastic sampling-based Bayesian inference, such as Markov chain Monte Carlo (MCMC) with gradient descent (GD), enhances multi-input multi-output (MIMO) detection by performing multiple GD steps per random walk for improved sampling efficiency, accelerated via Nesterov’s Accelerated Gradient (NAG-MCMC)\cite{zhou2023gradient}. Initially designed for QAM constellations in MIMO, its application to radar detection is indirect and less suited for single transmitter-receiver setups. Similarly, Principal Skewness Analysis (PSA) identifies directions with pronounced skewness for blind source separation in clutter suppression\cite{wang2023parallel}, while its moment-based variant improves target detection with computational efficiency~\cite{geng2015momentum,geng2014principal}. Another approach, Principal Kurtosis Analysis (PKA), extends the search for non-Gaussianity in remote sensing applications~\cite{meng2016principal}. Despite their differences, these methods do not leverage target scattering spread.

Deep learning-based radar detection methods~\cite{goodfellow2016deep} employ convolutional neural networks (CNNs), recurrent neural networks (RNNs), or transformer architectures on RD or RA maps, achieving significantly improved performance compared to classical signal processing techniques~\cite{wang2019study,brodeski2019deep}. Popular examples include PointNet and PointNet++, which operate on CFAR detections for multi-object detection and sea clutter classification by segmenting raw complex pulse echo data~\cite{chen2024pointnet,qi2017pointnet++}. Long Short-Term Memory (LSTM) based neural networks have been employed by treating scatterers as time sequences in fast time to discriminate closely spaced targets~\cite{kulkarni2023detection}. More recently, Kolmogorov–Arnold Network (KAN) based models have been applied for large target detection~\cite{kulkarni2025kan}, and transformer-based architectures have shown promise in extended target detection and tracking~\cite{guo2025jdtformer}. Despite their superior detection capabilities, these deep learning approaches are typically developed for controlled environments and suffer from limited explainability, lack of statistically interpretable models, and face significant challenges when deployed on the resource-constrained embedded hardware common in automotive radar systems.

 Detections from each dwell may exhibit either sparsity or high density in point clouds, depending on the allowable false alarm rate. Although dense point clouds are desirable for target characterization, increasing the number of CPI dwells to this cause increases complexity for their association into a target track~\cite{scheiner2019multi}. Additionally, high-density detections add complexity to resolving the target of interest, potentially necessitating more dwells. Multi-frame detection using Dynamic Programming (DP-MFD) is implemented in various military and commercial applications~\cite{yi2020multi} involving multiple dwells. Any innovative technique addressing target detection issues is expected to be characterized by low latency, a reduced number of dwells (preferably a single dwell), less complexity, and a large number of point clouds for target characterization~\cite {sun2020mimo}. 

This paper addresses the binary hypothesis detection problem for automotive extended targets by combining statistical machine learning and classical signal processing techniques, with an emphasis on leveraging statistical features from the scattering spread in the RD map while maintaining low computational complexity. An end-to-end detection pipeline is proposed to replace the conventional CFAR-based approach. The main contributions of this work are summarized as follows:

\begin{itemize}
\item Unlike traditional cell-based detection methods, we begin by highlighting the significance of RD segment-based detection for large targets. A binary hypothesis testing framework is established for the RD segment under examination, based on its statistical distribution.
\item To estimate the distribution parameters, algorithms based on Maximum Likelihood Estimation and Bayesian Gibbs Sampling are developed. Building on the latter approach, a simplified skewness-based test statistic is proposed for real-time target detection.
\item A revised detection pipeline is presented in which RD segment-under-detection is swept across the entire RD map. The redundant detections corresponding to extended targets are combined to minimize the target representation, thereby easing association and tracking.
\item Simulation and experimental studies present the performance of the proposed detection technique against OS-CFAR. The performance of the detection pipeline from frame to frame without association is shown on real data. The efficacy of the proposed test statistic in the presence of two closely-situated targets is also presented.
\end{itemize}
\section{Problem Statement}
Consider an ego vehicle equipped with a frequency-modulated continuous-wave (FMCW) radar operating at a frequency $f_0$ with chirprate $\mu$. $L$ chirps are transmitted with chirp repetition interval $T_{\mathrm{cri}}$ within one frame/coherent processing interval (CPI). At millimeter-wave frequencies, the wavelength $\lambda_0 <<$ target size, resulting in a range resolution that is much smaller than the target size. Consequently, the response from various parts of the vehicle are dispersed across multiple bins, while their orientation and relative motion with respect to the radar contribute to the observed Doppler spread. 

For the $k$th target with $I_K$ such scatterers, we attribute $(p_{k_i}, R_k\pm\delta R_{k_i}, v_k \pm \delta v_{k_i}), \forall i\in [1,I_k]$, where $p_{k_i}$ is the received power from the $i$th scatterer at a range $R_k\pm\delta R_{k_i}$ having relative radial velocity, $v_k\pm\delta v_{k_i}$. The down-converted signal received by the superheterodyne receiver is given by,
{\small
\begin{equation}
    \begin{split}
        y(t,l) &\approx \sum_{k=1}^K \sum_{i=1}^{I_k} \sqrt{p_{k_i}} \bigg(e^{\jmath 2\pi\big((f_{R_{k}}\pm \Delta f_{R_{k_i}})t+(f_{D_k} \pm \Delta{f_{D_{k_i}}})t_l+\phi_k\big)} \bigg) \\
        &\quad \quad \quad \quad \quad \quad + \eta(t,l).
        \label{eq:radar IF signal}
    \end{split}
\end{equation}
}
where $t_l=(l-1)T_{\mathrm{cri}}$ is the slow-time variable, $f_{R_k}\pm\Delta f_{R_{k_i}} =2\mu(R_k\pm\delta R_{k_i})/c $ and $f_{D_k}\pm\Delta f_{D_{k_i}}$ are the range and Doppler frequencies corresponding to the $i$th scatterer of the $k$th target, respectively, $\phi_k = \pi\mu\tau_k^2$ is a function of the two-way propagation delay, $\tau_k$, and $\eta(t,l)$ is the additive noise.

Applying the two-dimensional Discrete Fourier transform (2D-DFT) on~\eqref{eq:radar IF signal}, the resulting output is given by,
\begin{equation}
\begin{split}
Y(f_R,f_D) &= \sum_{k = 1}^{K} \sum_{i=1}^{I_k} \sqrt{p_{k_i}} \big( \delta(f_R - f_{R_k} - \Delta f_{R_{k_i}})  \\
& \delta(f_D - f_{D_k} - \Delta f_{D_{k_i}})\big) \quad + \eta(f_R,f_D).
\label{eq:root_RD-map}
\end{split}
\end{equation}
Here, $\delta(\cdot)$ denotes the Dirac delta function, which signifies the presence or absence of the target response. When the standard deviations of $\Delta f_{R_{k_i}}$ and $\Delta f_{D_{k_i}}$ across all scatterers exceed the range and Doppler resolutions, respectively, the target scatterer response disperses in the range and Doppler dimensions. This dispersion is clearly manifested in the RD map obtained after square-law detection,
\begin{equation}
    Z(f_R,f_D) = |Y(f_R,f_D)|^2.
    \label{eq:RD-map}
\end{equation}
This operation increases the signal-to-noise ratio (SNR), thereby improving radar target detection, which is posed as a binary hypothesis test.

Most often, detection is performed cell-by-cell across the RD map, where the test statistic compares the cell-under-test (CUT) against background noise estimated from neighboring cells. Guard cells are typically excluded during noise estimation to mitigate Fourier spectral leakage and accommodate variable target sizes. However, extended targets with a large spread can bias the noise estimate and hinder their overall detection, leaving some target RD cells undetected.

\begin{figure}[htp]
    \vspace{-1.9cm}
    \raggedleft
    \includegraphics[trim={1.5cm 4.8cm 3cm 8.0cm},width=8.85cm,height=10.0cm,]{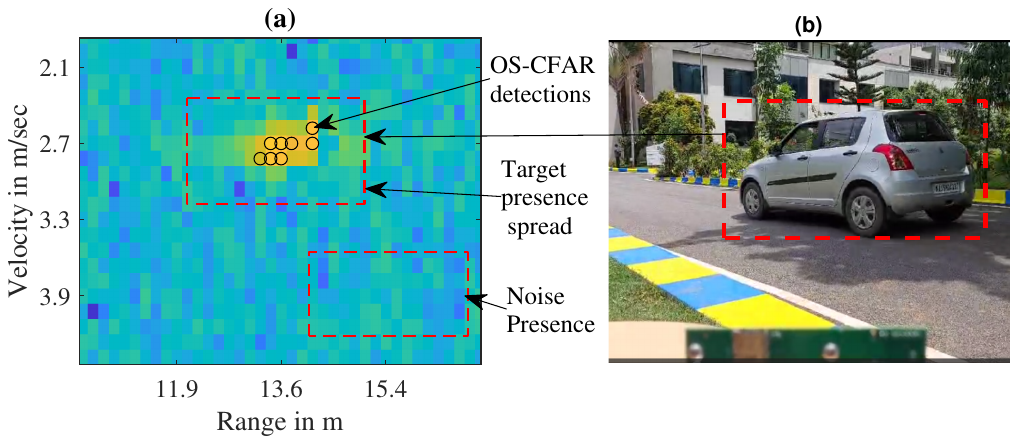}
    \vspace{-4.0cm}
    \caption{Experimental data for the vehicular target is shown in (b), with its corresponding RD map in (a).}
    \label{fig:tgt_spread_cfar}    
\end{figure}

To illustrate this, the Ordered-Statistic Constant False Alarm Rate (OS-CFAR) method, widely used in modern automotive mmWave radars, is applied on the data collected using a TI mmWave radar for the target shown in Fig.\ref{fig:tgt_spread_cfar}(b). The corresponding RD map in Fig.\ref{fig:tgt_spread_cfar}(a) exhibits spread across the range and Doppler dimensions, along with the detections obtained from OS-CFAR.

Although OS-CFAR is observed to provide more than one detection for the target, there are several RD cells with target response that go undetected. Furthermore, the point clouds obtained from these detections are subjected to association using Probabilistic Data Association (PDA) or its variant, Joint PDA (JPDA)~\cite{fortmann1983sonar,fortmann1980multi,rezatofighi2015joint}. The real-time computational expense for the association of such a target with multiple number of detections increases significantly. Given these limitations, the problem at hand is to accurately delineate the full spatial extent of targets' scattering profiles within the RD map. 

\section{Proposed Approach}
To capture the spatial extent of a target’s scattering profile, a localized region within the RD map, termed the RD segment, is defined. Rather than a conventional cell-by-cell search, the RD segment is systematically slid across the RD map to detect extended targets. A detection technique is then applied to the RD segment as a whole, rather than on individual cells. In the following section, a detailed formulation of this RD segment-based detection approach is presented.

\subsection{Binary Hypothesis Construction for an RD segment}
\begin{figure}[htbp]
\centering
    \vspace{-0.3cm}
    \includegraphics[trim={0.5cm 0 3.5cm 0},width=0.485 \textwidth, height=0.28\textwidth]{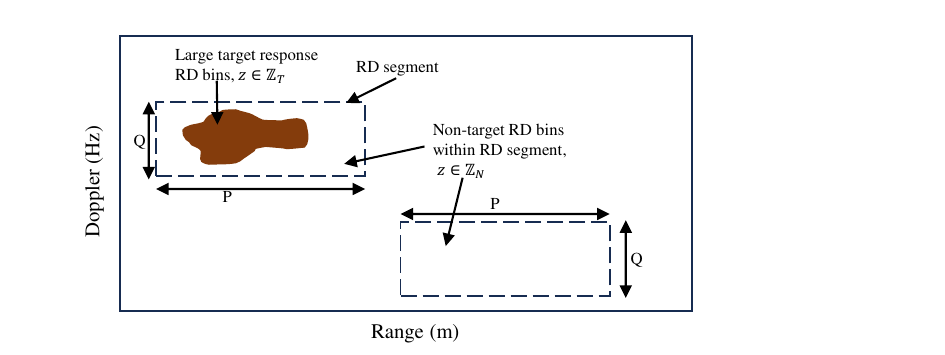}    
    \vspace{-0.8cm}
    \caption{Illustration of localized RD segment in RD map.}
    \label{fig:Illustration_of_localized_map}
\end{figure}

The dimensions of the RD segment is intended to encapsulate the extended target responses in the RD map, sized to cover typical automotive vehicles (e.g., cars, SUVs, mini trucks) with spatial extents of $4-6~$m in size. Figure.~\ref{fig:Illustration_of_localized_map} illustrates one RD segment containing a target, while another containing none. Let $\mathbb{Z_T}$ and $\mathbb{Z_N}$ represent the sets of RD cells corresponding to target and noise responses, respectively. The first segment includes cells from both $\mathbb{Z_T}$ and $\mathbb{Z_N}$, indicating target presence, whereas the second contains cells only from $\mathbb{Z_N}$, showing absence of a target. Each RD segment spans $P$ range bins ($6$ m) and $Q$ Doppler bins ($2$ m/s). Unlike traditional cell-based methods, the proposed approach treats the entire RD segment as the detection region, ensuring the full target response is captured. Recognizing the possibility that responses from multiple targets, such as closely-spaced ones, may coexist within a single RD segment, the aim is to detect RD segments with target responses, irrespective of the number of targets present.

Let the values of the RD cells $(\tilde{p},\tilde{q}), \tilde{p}\in [1,P], \tilde{q}\in[1,Q]$ within an RD segment (size:$P\times Q$) be denoted as $z_{\tilde{p},\tilde{q}}\in\mathbb{R}$. The binary hypothesis test is then formulated as
\begin{align}
\begin{split}
    H_0 &: {z_{\tilde{p},\tilde{q}} \in \mathbb{Z_N}} \\
    H_1 &: {z_{\tilde{p},\tilde{q}} \in \mathbb{Z_T} \cup \mathbb{Z_N}},
    \label{eq:statistical_test_over_rdseg}
\end{split}
\end{align}
where $H_0$ and $H_1$ denote the null and alternate hypotheses, respectively. As illustrated in Fig.~\ref{fig:Illustration_of_localized_map}, the sets $\mathbb{Z_T}$ and $\mathbb{Z_N}$ correspond to the target scattering spread and noise, within an RD segment. Ideally, only the segment encompassing $\mathbb{Z_T}$ is of interest, as the target scattering is confined to this region. However, in real-world automotive scenarios, the scattering spread for a target varies with  target characteristics, environmental conditions, and incidence angles. Due to such variations in target response, we perform detection over predetermined RD segment dimensions ( ($6$m,$2$m/s), without loss of generality) that capture most of the target scattering response along with limited interference noise. Consequently, the alternative hypothesis $H_1$ is defined as the union of $\mathbb{Z_T}$ and $\mathbb{Z_N}$. The probability density function (pdf) of the observations $z_{\tilde{p},\tilde{q}}$ under the two hypotheses can be written as,

{\small
\begin{equation}
 f_z(z)=\begin{cases}
    f_z(z_{\tilde{p},\tilde{q}} \in \mathbb{Z_N}\mid H_0) \sim f_{z_{\tilde{p},\tilde{q}}}(|\eta(f_{R_{\tilde{p}}},f_{D_{\tilde{q}}})|^2),      \\
    f_z(z_{\tilde{p},\tilde{q}} \in (\mathbb{Z_T} \cup \mathbb{Z_N}) \mid H_1)  \sim f_{z_{\tilde{p},\tilde{q}}}(|Y(f_{R_{\tilde{p}}},f_{D_{\tilde{q}}})|^2) .
  \end{cases}
  \label{eq:signal_model}
\end{equation}
}

The hypotheses $H_0$ and $H_1$ correspond to the pdf of the square-law detector outputs of $\eta(f_R,f_D)$ and $Y(f_R,f_D)$, respectively. Typically, radar systems encounter interference noise $\mathbb{Z_N}$ due to various environmental conditions such as clear atmospheric weather, fog, rain, snow, or mixed scenarios like rain combined with fog. The statistical characteristics of these noise sources are commonly modeled using Gaussian, Weibull/log-normal/K-distributions, Rayleigh, and Gaussian mixture models~\cite{skolnik1980introduction,skolnik2008radar,richards2014fundamentals,doviak1994doppler,liebe1989millimeter}. In contrast, target reflections $\mathbb{Z_T}$ are modeled using Swerling cases 1 through 4, where the distributions of scattered power range from exponential to chi-square with four degrees of freedom~\cite{skolnik1980introduction,skolnik2008radar,richards2014fundamentals,richards2010principles}.
\subsubsection{Noise and Target response distributions}
In this paper, we consider \( \mathbb{Z_N}\) to follow additive white Gaussian noise (AWGN). The extended target is modeled using the Swerling-3 model, wherein the power of the scatterers follows a Chi-squared distribution with four degrees of freedom, \( \chi^2_4 \). In the  presence and absence of a target response, the scatterer power ($p_{k_i}$) in~\eqref{eq:root_RD-map} and the noise are modeled to follow, 
\begin{align}
&p_{k_i}\sim 
\begin{cases}
\delta_{0},   & \text{under }H_{0}\\
\chi^{2}_{4}, & \text{under }H_{1}\\
\end{cases} \nonumber \\
 &\eta(t,l)\sim \mathcal{CN}(0,\sigma^2), 
\label{eq:Pdf_models_time-domain}
\end{align}
where $\delta_0$ is point mass at zero and $\sigma$ is the standard deviation of complex AWGN noise. 
The data under $H_0$ follows \( f_{z}(z \mid H_0) \), after square-law detection follows an exponential distribution—or equivalently, a chi-squared distribution with two degrees of freedom, \( \chi^2_2 \)~\cite{Papoulis2002probability,bertsekas2008introduction}—which is a special case of the Gamma distribution. The rate of decay of this distribution is governed by \( 1/\sigma \).

The cells within the RD segment are assumed to be independent and identically distributed (i.i.d.). Therefore, under $H_0$, the joint pdf of all the cells is given by,
\begin{align}    
        f_z(z_{1,1},z_{1,2},\dots,z_{\tilde{p},\tilde{q}} \mid H0) \simeq&\prod_{\tilde{p},\tilde{q} \in \mathbb{Z_N}}e^{(-\sigma^2 \frac{z_{\tilde{p},\tilde{q}}}{2})}.
    \label{eq:bimodal_noise}
\end{align}
Under the alternate hypothesis in contrast, the presence of some cells belonging to $\mathbb{Z}_{\mathbb{T}}$ and the others belonging to $\mathbb{Z_N}$ results in the distribution $f_z(z\mid H_1)$ not yielding a closed-form expression. Depending on the target's characteristics, the resulting distribution may either resemble or deviate significantly from the noise distribution. Consequently, the distribution over the RD segment—corresponding to the combined region $\mathbb{Z_T}\cup \mathbb{Z_N}$, exhibits a bimodal nature. To capture this behavior, \( f_{z}(z \mid H_1) \) is modeled as a weighted mixture of Gamma distributions, expressed as

\begin{align}    
         \hspace{0pt} f_z(z_{1,1}, z_{1,2}, \dots, z_{\tilde{p},\tilde{q}} \mid H_1) \simeq& \prod_{\tilde{p},\tilde{q} \in \mathbb{(Z_T  \cup  Z_N)}} \nonumber \\\hspace{0pt} \Bigg(W_1\frac{\beta_1 ^{\alpha_1}}{\Gamma(\alpha_1)} z_{\tilde{p},\tilde{q}}^{\alpha_1 - 1}e^{(-\beta_1 z_{\tilde{p},\tilde{q}})}  &+    W_2\frac{\beta_2 ^{\alpha_2}}{\Gamma(\alpha_2)} z_{\tilde{p},\tilde{q}}^{\alpha_2 - 1}e^{(-\beta_2 z_{\tilde{p},\tilde{q}})}\Bigg) ,        
    \label{eq:bimodal_wtd_gamma}
\end{align}
where $W_j$, $\alpha_j>0$, $\beta_j>0$, $j\in \{1,2\}$ are the weights, shape and rate parameters of Gamma distributions. For bimodal Gamma mixture model, $W_1+W_2 = 1$. This representation gives joint distribution of the RD segment, comprising i.i.d. target scatterers and noise.

\subsubsection{Test statistic}
Having modeled the probability distributions of the RD segment scatterers for $H_0$ and $H_1$, the test statistic for the binary hypothesis is given by,

\begin{equation}
    \mathbb{F}\big(f_Z(z;\theta)\big)
    \;\underset{H_0}{\overset{H_1}{\gtrless}}\; T.
    \label{eq:binary_hypothesis_ratio}
\end{equation}
In this formulation, \( \theta \in \{W_j, \alpha_j, \beta_j\} \) represents the set of model parameters used to characterize the statistical behavior under the hypotheses \( H_0 \) and \( H_1 \). The function \( \mathbb{F}(\cdot) \), applied to the probability density function \( f_Z(z;\theta) \) of the received RD segment, yields a scalar test statistic that is compared against a threshold \( T \) to perform binary hypothesis testing.

Under hypothesis $H_0$, the distribution is modeled as an exponential, which is equivalent to a Gamma distribution with shape parameter $\alpha = 1$ and a rate parameter $\beta$ determined by the signal-to-noise ratio (SNR). Under hypothesis $H_1$, the parameters $\theta$ will be estimated using Maximum Likelihood Estimation (MLE) and Gibbs sampling, as described in the following subsection. These parameter estimates are then used to construct the decision function \(\mathbb{F}(\cdot) \), which facilitates effective discrimination between the two hypotheses.

\subsection{Parameter estimation}
Maximum Likelihood Estimation (MLE) is performed by minimizing the negative log-likelihood (NLL) to obtain point estimates of the model parameters under the alternative hypothesis \( H_1 \). This process also facilitates analysis of the likelihood surface to determine whether the underlying distribution is unimodal or bimodal—an insight that is crucial for guiding posterior inference in a Bayesian framework.

\subsubsection{Likelihood Formulation for Gamma Mixture Model}
Maximizing the likelihood of the Gamma mixture model in~\eqref{eq:bimodal_wtd_gamma} is equivalent to minimizing its negative log-likelihood (NLL) over the distribution of the RD segment under $H_1$ comprising \( \mathbb{N} = P \times Q \) cells, expressed as,
\begin{align}
    & \min_{\theta} \text{NLL}(\theta)  \quad \text{where,}\nonumber \\
    \text{NLL}(\theta) &=-\mathcal{L}(z;\alpha,\beta) = -log(f_z(z \mid H_1)) \nonumber\\    
    \text{NLL}(\theta) &= \sum_{j=1}^{\mathbb{N}} -log\bigg (W_1\frac{\beta_1 ^{\alpha_1}}{\Gamma(\alpha_1)} z_{j}^{\alpha_1 - 1} \exp{(-\beta_1 z_{j})} \nonumber \\ 
        &+  W_2\frac{\beta_2 ^{\alpha_2}}{\Gamma(\alpha_2)} z_{j}^{\alpha_2 - 1} \exp{(-\beta_2 z_{j})} \bigg).        
        \label{eq:NLL}
\end{align}
Since a closed-form solution for the parameters under hypothesis \( H_1 \) is intractable from~\eqref{eq:NLL}, the estimation is carried out using gradient-based optimization across multiple batches of $H_1$ RD segments. To promote stable convergence during this process, all RD segments undergo global normalization~\cite{cabello2023impact}. This normalization step accelerates convergence and prevents issues such as exploding or vanishing gradients, which may arise due to signal strength variations caused by distance-dependent attenuation. The following Algorithm~\ref{algo:Gradient point estimation} outlines the proposed procedure for estimating the optimal parameters.
\begin{algorithm}
{\small
\caption{MLE-Based Gradient Optimization for parameter estimation of $f(z;\theta |H_1)$}
\begin{algorithmic}[0] 
    
    \State Initialize parameters $\theta = \begin{bmatrix} W_1,\alpha_1,\beta_1,W_2,\alpha_2, \beta_2 \end{bmatrix}^T$
    \State Set batches $\mathbb{B}$, per-segment bins $\mathbb{N} = P \times Q$ \Comment{$H_1$ RD segments}    
    \State Set learning rate $\tilde{\eta}$ and initial loss NLL$^{(0)}(\theta)=0$
    \For{$t = 1$ to $N_{iter}$}
    \State Update $f_z(z\mid \theta)$ as in~\eqref{eq:bimodal_wtd_gamma} \Comment{with $\begin{bmatrix} W_1,\alpha_1,\beta_1,W_2,\alpha_2, \beta_2 \end{bmatrix}$ }
        \State NLL$^{(t)}(\theta)=\frac{1}{\mathbb{N}\mathbb{B}}\sum_{j=1}^{\mathbb{B}}$NLL($\theta$) \Comment{as in ~\eqref{eq:NLL}}
        \If{$|\text{NLL}^{(t)}(\theta) - \text{NLL}^{(t-1)}(\theta)| < \text{threshold}$}
            \State \textbf{break} \Comment{Exit loop, reached local or global minima}
        \EndIf
        \State Compute the gradients $\nabla \text{NLL}^{(t)}(\theta)$ 
        \State Update parameters: $\theta \gets \theta - \tilde{\eta} \times  \nabla \mathbb{\text{NLL}}^{(t)}(\theta)$  
        \State Updated parameters $\begin{bmatrix} W_1,\alpha_1,\beta_1,W_2,\alpha_2, \beta_2 \end{bmatrix} = \theta$
        \State NLL$^{(t-1)}(\theta)$= NLL$^{(t)}(\theta)$
    \EndFor
    \State Return optimized parameters $\theta^*$
\end{algorithmic}
\label{algo:Gradient point estimation}
}
\end{algorithm}

\begin{figure}[htbp]    
    \centering
    \includegraphics[trim={0.5cm 0.5cm 0cm 0.3cm},width=8.8cm,height=5cm,]{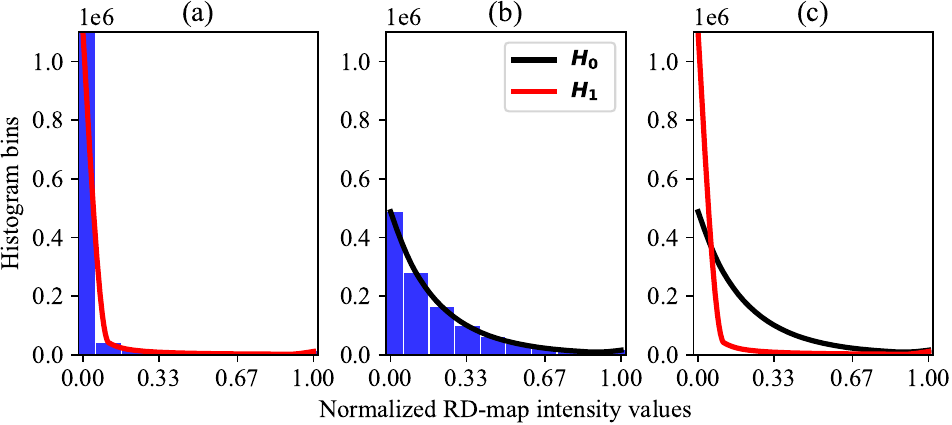}
    \vspace{0cm}
    \caption{Histograms for 10500 samples: (a) Alternate hypothesis, (b) Null hypothesis, (c) envelopes for the two hypothesis.}
    \label{fig: Binary hypothesis}
    \vspace{-0.5cm}
\end{figure}

Based on the Swerling-3 model for extended targets with AWGN, as defined in~\eqref{eq:Pdf_models_time-domain}, Section~V reports the MLE results over approximately $10,500$ RD segments under $H_1$. The estimation converged to a nearly unimodal Gamma mixture with parameters:
\[
\begin{aligned}
W_1 &= 0.005,\quad \alpha_1 = 0.131,\quad \beta_1 = 35.1, \\
W_2 &= 0.995,\quad \alpha_2 = 0.132,\quad \beta_2 = 10772.1.
\end{aligned}
\]
Despite the RD segments under the alternate hypothesis being drawn from \( \mathbb{Z_T} \cup \mathbb{Z_N} \), the data is effectively captured by a single Gamma component with \( W_2 \simeq 99.5\%\). This observation is supported by Fig.~\ref{fig: Binary hypothesis}(a), which shows the histogram of RD segments used in the MLE procedure. The empirical distribution exhibits a distinctly unimodal shape, closely matching a single Gamma distribution and confirming the suitability of a unimodal fit. For reference, Fig.~\ref{fig: Binary hypothesis}(b) presents the histogram of the RD segments under the null hypothesis (no targets), along with its envelope. The noise-only data conforms to an exponential distribution, a special case of the Gamma distribution with shape parameter \( \alpha = 1 \).
Although the data from the two hypotheses follow unimodal Gamma distributions, Fig.~\ref{fig: Binary hypothesis}(c) shows the contrast in their distributions, which has to be exploited for an appropriate decision.

It is to be emphasized that the Gamma mixture model is simplified to a single Gamma distribution with parameters \( \theta \in \{\alpha, \beta\} \)when $\mathbb{Z}_{{\mathbb{N}}}$ is drawn from AWGN. For different scenarios, however, the noise is modeled using Weibull, K-distributed, or log-normal distributions. The pdf of the RD segments with Swerling targets in such cases will have different weights $W_1$ and $W_2$, requiring appropriate treatment.

\subsubsection{Bayesian inference using MCMC Gibbs sampling}
While MLE provides a point estimate, i.e., a single best-fit value of the model parameters based on observed data, it does not adequately capture the uncertainty and variability inherent in the data~\cite{gelman1995bayesian,lambertstudent}. To address this limitation, Bayesian inference is adopted with Markov Chain Monte Carlo (MCMC) sampling, which enables full posterior inference over the parameter space. This approach is particularly beneficial in distinguishing hypothesis $H_1$ from $H_0$, where the distributions, as illustrated in Fig.~\ref{fig: Binary hypothesis}(c), exhibit considerable overlap and test statistic demanding a threshold $T$ derived from parameters ($\mathbb{F}(\theta)$ as in \eqref{eq:binary_hypothesis_ratio}). Under such settings, MCMC provides a principled framework to capture parameter uncertainty and characterize the underlying distribution more effectively.

MCMC sampling is grounded in Bayes’ theorem, which provides a framework for updating our beliefs about the model parameters $\theta$ given the observed data from the target RD segment. Gibbs sampling, an MCMC-based method, is attractive for approximating the posterior distribution $f_z(\theta\mid z)$, especially when the analytical solution is intractable. Gibbs sampling iteratively samples from the conditional distributions of each parameter, efficiently exploring the posterior space~\cite{gelman1995bayesian,murphy2022probabilistic,lambertstudent,geman1984stochastic}.  In each iteration, $t$, of the Gibbs sampler cycles through the two parameters $(\tilde{\alpha},\tilde{\beta}) \in \theta$ of Gamma distribution, drawing $\tilde{\beta}$ conditioned on the value of $\tilde{\alpha}$ and vice versa, i.e,
\begin{align} 
    f_z(\tilde{\beta}^{(t)} &\mid \tilde{\alpha}^{(t-1)}, z)\\
    f_z(\tilde{\alpha}^{(t)}  &\mid \tilde{\beta}^{(t)}, z).
\end{align}
$\tilde{\beta}$ is drawn from the posterior Gamma distribution (as derived in Appendix B~\eqref{eq:posterior_for_gibbs}),
\begin{align}
  f_z(\tilde{\beta}^{(t)} \mid (z,\tilde{\alpha}^{(t-1)})) \sim &~ \text{Gamma}(a+\mathbb{N}\tilde{\alpha}^{(t-1)},b+S_z),
    \label{eq:beta_sampling}
\end{align}
where $S_z = \sum_{j=1}^{\mathbb{N}} z_j$, $a,b$ are non-informative prior parameters set to zero, the posterior distribution for $\tilde{\beta}$ simplifies to:
\begin{align}    
    f_z(\tilde{\beta}^{(t)} \mid z) &\sim \text{Gamma}(\mathbb{N}\tilde{\alpha}^{(t-1)},S_z). 
    \label{eq:pos_rate}
\end{align}
With $\tilde{\beta}^{(t)}$ obtained, $\tilde{\alpha}^{(t)}$ is estimated using the Newton-Raphson method~\cite{chong2013introduction,bertsekas1982projected,kochenderfer2019algorithms}, employing the gradient and curvature of the posterior surface as derived in Appendix~A. Algorithm~\ref{algo:Gibbs Newton} outlines the steps to estimate both the parameters of the Gamma distribution under $H_1$ with the training data used for MLE earlier. To ensure that parameter updates follow the descent direction and to improve numerical stability, the Levenberg-Marquardt modification~\cite{chong2013introduction} is incorporated, using a damping factor of $\epsilon = 10^{-6}$.
\begin{algorithm}
{\small 
\caption{Newton method for posterior estimation using Gibbs sampling}
\begin{algorithmic}[0] 
    \State Set RD segment bins $\mathbb{D}=P \times Q $ \Comment{Data dimension}
    \State Set Number of $H_1$ RD segments as, $\mathbb{B}$  
    \State $\mathbb{N} = \mathbb{B} \times \mathbb{D}$ \Comment{Total number of RD segment cells}
    \State ${S_z} =\sum _{j=1}^{\mathbb{B}}\sum_{k=1}^{\mathbb{D}} z_k^{(j)}$
    \State ${LS_z} ={\sum _{j=1}^{\mathbb{B}}\sum_{k=1}^{\mathbb{D}} \log(z_k^{(j)})}$
    \State $t=0$, Initialize parameters $\theta = \begin{bmatrix} \tilde{\alpha}_0,\tilde{\beta}_0\end{bmatrix}^T$ , $\epsilon=10^{-6}$
    \State $\alpha_c = \tilde{\alpha}_0; \beta_c = \tilde{\beta}_0$ 
    \For{{$t = 1$ to $N_{iter}$} , $z \in $ \{$H_1$ RD segments\} }
        \State $\tilde{\beta}^{(t)} \sim \text{Gamma}(\mathbb{N}\tilde{\alpha}^{(t-1)},S_z)$ \Comment{Gibbs sample from ~\eqref{eq:pos_rate}}
        \State $\tilde{\alpha}^{(t)} = \tilde{\alpha}^{(t-1)}$ 
        \For{{$u = 1$ to U$_{iter}$}}  \Comment{test convergence}\\
\Comment{$g$ and $h$ expressions given in Appendix A}
           \State $g =\mathbb{N}(\log(\tilde{\beta}^{(t)}) - \Psi(\tilde{\alpha}^{(t)})) + LS_z$ \Comment{$\frac{\partial(\mathcal{L})}{\partial\tilde{\alpha}}$, ~\eqref{eq:grad_alpha}}
            \State $h = -\mathbb{N}\Psi'(\tilde{\alpha}^{(t)})$ \Comment{$\frac{\partial^2(\mathcal{L})}{\partial\tilde{\alpha}^2}$,~\eqref{eq:Hess_alpha}}\\
            \State $\tilde{\alpha}^{(t)} = \tilde{\alpha}^{(t)} - \frac{g}{(h+ \epsilon)}$ \Comment{Newton-Raphson method}
            \State $\tilde{\alpha}^{(t)} = max(\tilde{\alpha}^{(t)},\epsilon)$
            \If{$| \frac{g}{(h+ \epsilon)}| < \epsilon $}
                \State \textbf{break} \Comment{$\tilde{\alpha}$ convergence reached for sampled $\tilde{\beta}$}
            \EndIf            
        \EndFor
        \State $\alpha_c = \alpha_c \cup \tilde{\alpha}^{(t)} ;\beta_c = \beta_c \cup \tilde{\beta}^{(t)} $ \Comment{Store converged $\alpha,\beta$}
    \EndFor
    \State \(\bar{\alpha_c} = \mathbb{E}[\alpha_c] ,\quad \quad \bar{\beta_c} = \mathbb{E}[\beta_c]\) \Comment{Mean of the converged shape and rate parameters}
    \State Return optimized parameters $\theta^* = (\bar{\alpha_c},\bar{\beta_c})$
\end{algorithmic}
\label{algo:Gibbs Newton}
}
\end{algorithm}

The results of Algorithm~\ref{algo:Gibbs Newton} are detailed in Section~V, with some observations briefly highlighted here for context. For the set of $10,500$ RD segments  the estimated parameters are $\bar{\alpha}_c \simeq 0.13$ and $\bar{\beta}_c = 7682$. The rate parameter $\bar{\beta}_c$ is sensitive to target range, radar cross-section (RCS), and in-segment noise due to SNR variations governed by the radar range equation (RRE), which affects the scaling of RD segments at different distances; As derived in Appendix C, the rate parameter $\beta$ exhibits sensitivity to scaling, resulting in distinct $\beta$ values for global versus local normalization~\eqref{eq:normalization_beta}. Conversely, the shape parameter $\alpha$ demonstrates consistency across different scenarios, highlighting its robustness to SNR variations~\eqref{eq:normalization_alpha}.

The above analysis provides a quantitative estimate of the shape parameter \( \bar{\alpha}_c \) in the presence of a target. Building on this, the next section introduces a mechanism to estimate the detection threshold \( T \) in real time, based solely on the shape parameter \( \alpha \) of the RD segment under test.
\subsubsection{Statistical third-order moment function}
MLE and Bayesian inference are used to estimate the Gamma distribution parameters for modeling the pdf under $H_1$.. However, real-time detection based on the learned Gamma pdf is computationally demanding. The first two moments of the Gamma distribution depend on both the shape ($\alpha$) and rate ($\beta$) parameters. However, since $\beta$ is sensitive to normalization (Appendix~C), these moments are not reliable for robust detection. In contrast, the third central moment (skewness) depends only on $\alpha$, which is largely unaffected by normalization and thus serves as a reliable discriminator between the hypotheses. Accordingly, this work adopts skewness for real-time target detection, given by:
\begin{align}
    \mathbb{F}(f(z;\theta)) \propto \kappa = \mathbb{E}\left[\frac{(z - \mu_{\text{rdseg}})^3}{\sigma_{\text{rdseg}}^3}\right] = \frac{2}{\sqrt{\alpha}},
    \label{eq:skewness}
\end{align}
where \( \mu_{\text{rdseg}} \) and \( \sigma_{\text{rdseg}} \) are the mean and standard deviation of the RD segment, respectively. This sole dependency on $\alpha$ motivates the use of skewness as a criterion for distinguishing target-present RD segments from noise-only segments.

For an RD segment under test, the detection decision can be formulated as:
\begin{align}
    \mathbb{F}\big(f_z(z;\theta \in \alpha)\big) \simeq {\kappa} \;\underset{H_0}{\overset{H_1}{\gtrless}}\; T.
    \label{eq:skewness_threshold}
\end{align}
A large target response within the RD segment results in a low shape parameter, thereby yielding a large skewness value.
\subsection{RSP detection pipeline}
\begin{figure}[htp]
    \centering       
    \includegraphics[trim={0.5cm 0.5cm 0.5cm 0.75cm},width=0.489\textwidth]{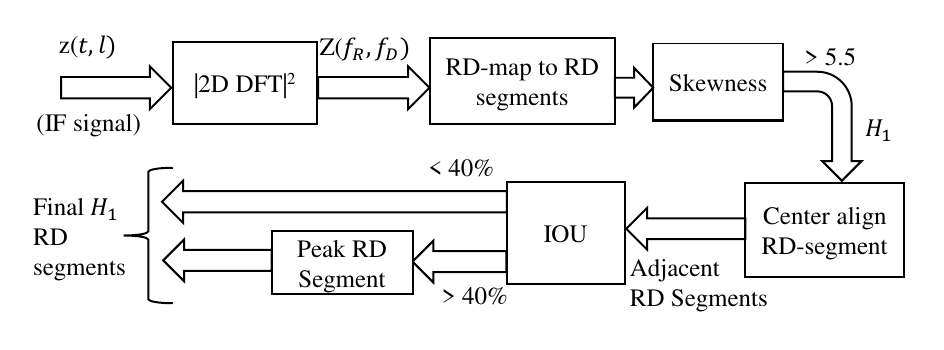}
    
    \caption{The proposed detection pipeline for RD segment-based detection.}
    \label{fig:hypothesis_separation}
\end{figure}

The proposed skewness-based detector operates on sliding RD segments across the entire RD map. In addition to detecting the segment containing the full target response, adjacent segments may also be flagged due to partial target presence, necessitating appropriate handling within the detection pipeline. The complete detection framework is illustrated in Fig.~\ref{fig:hypothesis_separation}.
The pipeline eliminates redundant segments and recenters each RD segment to align the peak intensity at its center, for extended target ($H_1$) segments. This centering is rooted in the dual DFT operations described in~\eqref{eq:RD-map}, where the scattering pattern exhibits a central peak at the target’s nominal location, with energy decaying radially due to sidelobe effects~\cite{nuttall1981some}. 

By aligning segments to center the peak response, the method effectively merges overlapping neighboring RD segments, thereby reducing false alarms and redundancy. This alignment requires evaluating adjacent segments for overlap, which is quantified using the Intersection over Union (IoU) metric~\cite{yu2016unitbox} to accurately represent target presence. An overlap exceeding 40\% is considered redundant; in such cases, the segment with the highest peak is retained, while segments with IoU below 40\% are preserved as distinct detection segment.

The proposed pipeline reduces redundancy in detected RD segments, aligning their count more closely with the actual number of targets. Unlike point-wise association in a point cloud, segment-level association across dwells becomes significantly simpler and more computationally efficient, making it well-suited for real-time implementation.

\section{Data Synthesis}

Table~\ref{tab:spec} provides the parameters chosen for the FMCW radar and the target model for synthesizing data. To ensure consistency, real experiments were conducted with the same radar parameters. A $4~$m long hatchback car is used for real data acquisition, encompassing various scenarios of approaching and moving away from the radar. The RD segment is defined based on the specifications outlined in Table~\ref{tab:Realworld_spec}.

\begin{table}[htbp]
\raggedleft
\caption{FMCW radar for simulation and real data}
\label{tab:simulated_data}
\begin{tabular}{|l|l|}
\hline
\textbf{Parameter} & \textbf{Value} \\ \hline
$f_0$& $77$GHz\\ \hline
$\mu$(slope)&$16.67$MHz/$\mu$sec \\ \hline
$T_{cri}$&$50 \mu$sec \\ \hline
$f_s$&$10$MHz \\ \hline
Fast Time samples&$256$ \\ \hline
Chirps & $128$ \\ \hline
Chirp type & Up chirp \\ \hline
\multicolumn{2}{|c|}{\textbf{Simulated Data}} \\ \hline
SNR Range & $-25$dB to $25$dB \\ \hline
Target Model & Swerling-3 ($\chi_4^2$ distribution) \\ \hline
Noise Type & AWGN \\ \hline
RCS (Side Views) & $19$ to $22~\mathrm{dBm}^2$ \\ \hline
RCS (Front Views) & $8.7$ to $20.5~\mathrm{dBm}^2$ \\ \hline
RCS (Rear Views) & $14.4$ to $24.6~\mathrm{dBm}^2$ \\ \hline
Target distances (m) & $15$ to $65$ \\ \hline
Target RD segments (MC trials) & 10500 \\ \hline
\multicolumn{2}{|c|}{\textbf{Real Data}} \\ \hline
Device & TI AWR2243 \\ \hline
Data capture Card & DCA1000 \\ \hline
Hatchback Car(Length x Width x Height) &$3.86$m $\times 1.735$m $\times 1.52$m \\ \hline
\end{tabular}
\label{tab:spec}
\end{table}

\begin{table}[!ht]
\centering
\caption{RD Segment Specifications}
\small 
\setlength{\tabcolsep}{4pt} 
\renewcommand{\arraystretch}{1.1} 
\begin{tabular}{|c|c|c|c|}
\hline
\textbf{Range Res.} & \textbf{Velocity Res.} & \textbf{Dimensions} & \textbf{RD bins} \\ \hline
$0.3516$\,m & $0.3044$\,m/s & $5.98$\,m $\times$ $2.13$\,m/s & $17 \times 7$ \\ \hline
\end{tabular}
\vspace{-2mm}
\label{tab:Realworld_spec}
\end{table}

Extended targets are simulated using $50$–$100$ scatterers drawn from a $\chi^2_4$ power distribution. The target scattering spread is modeled by randomly placing scatterers within a $\pm1.6$,m vicinity of the target's nominal position, simulating side, rear, or front-facing perspectives. The corresponding radar cross sections (RCS) are detailed in Table~\ref{tab:simulated_data}.
In velocity, the spread is limited to $\pm1.065$\,m/s relative to the target’s motion. Additionally, a uniform spread of $0.3$\,m/s per dwell is introduced to generate realistic Doppler diversity in the extended target profile. RD maps are obtained by applying a windowed 2D-DFT to the raw radar signals.
RD segments of size $5.98$\,m $\times$ $2.13$\,m/s are extracted by sliding overlapping windows across the RD map. With known ground truth, these segments are classified into target ($H_1$) and noise ($H_0$) categories. A subset is used for Gamma distribution parameter estimation via MLE and Bayesian inference, while the complete dataset is utilized to evaluate the skewness-based detection pipeline.

RD-segment data collected from both simulated and real-world sources—primarily under target-present ($H_1$) conditions—are used to estimate the pdf parameters for the detection model. To ensure consistent scaling across all RD segments and enable stable gradient-based optimization, global normalization~\cite{ioffe2015batch,cai2019quantitative} is applied. This involves selecting the maximum intensity value from the training set of $H_1$ RD segments and using it to normalize each segment and its constituent RD bins.

The following section presents the estimation of Gamma distribution parameters using Maximum Likelihood Estimation (MLE) and Bayesian inference via Gibbs sampling, applied to training samples of simulated RD segments under the alternate hypothesis ($H_1$). Subsequently, the detection pipeline is evaluated using both simulated and real-world datasets to demonstrate its performance in terms of detection accuracy and robustness.
\section{Results and Analysis}
\subsection{Optimization results}
\begin{figure*}[!t]
  \centering
  \includegraphics{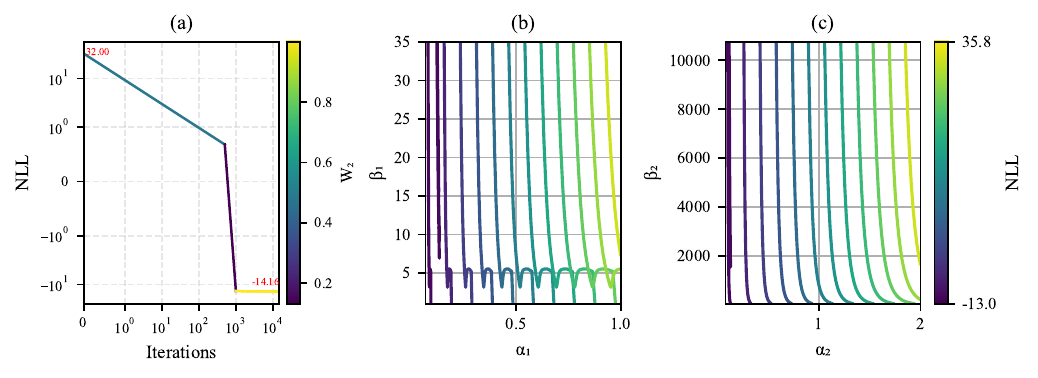}
  \caption{MLE for 2 components of Gamma mixtures: (a) NLL optimization with $W_2$ colorbar\;; Contour plots of the parameter surfaces for (b) $\text{Gamma}_1(W_1,\alpha_1,\beta_1)$ and (c) $\text{Gamma}_2(W_2,\alpha_2,\beta_2)$, shown with a shared NLL colorbar.}  
  \label{fig:Gamma_mix_contours}  
\end{figure*}

Maximum likelihood parameter estimation, as outlined in Section~{III}(B), is first performed on a dataset comprising $10{,}500$ RD segments under the alternate hypothesis ($H_1$). The bimodal Gamma mixture model parameters are estimated via a gradient-based optimization procedure (Algorithm~\ref{algo:Gradient point estimation}), which minimizes the negative log-likelihood function. The convergence trajectory of the NLL, together with the contour plots of the estimated parameters, is depicted in Fig.~\ref{fig:Gamma_mix_contours}, thereby illustrating both the convergence behavior and the parameter interaction landscape.

In Fig.~\ref{fig:Gamma_mix_contours}(a), the NLL is observed to converge to $-14.16$ dB over $10{,}000$ iterations. Figs.~\ref{fig:Gamma_mix_contours}(b) and (c) present the shape and rate parameter surfaces with NLL loss contours, capturing the behavior of individual components. At convergence, the optimization yields $(\alpha_1,\beta_1)=(0.131,35.1)$ and $(\alpha_2,\beta_2)=(0.132,10772.1)$ with weights $W_1=0.5\%$ and $W_2=99.5\%$. With $W_2$ completely representing the dataset, one can conclude that a single Gamma component is sufficient to model the data under $H_1$ ($\theta={\alpha_2,\beta_2}$).

\begin{figure}[H]
    \centering
    \includegraphics[trim={0.5cm 0.5cm 0cm 0.25cm},width=0.485\textwidth]{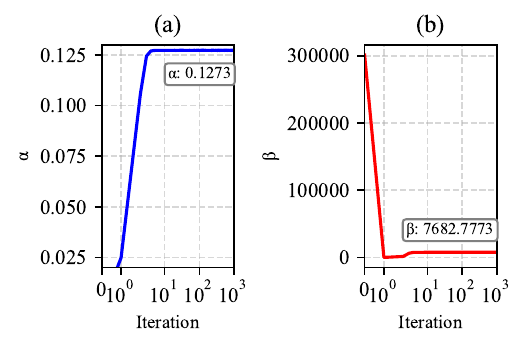}
    \caption{Gibbs Sampling posterior estimation  (a) $\alpha$ estimation (b) $\beta$ estimation.}
    \label{fig:Gibbs_posterior_estimation}
\end{figure}

After concluding that the distribution is unimodal, the posterior surface of the Gamma distribution is explored using Gibbs sampling for $H_1$ RD segments, as outlined in Algorithm~\ref{algo:Gibbs Newton}. For the dataset used earlier for MLE, the convergence of the shape ($\alpha$) and rate ($\beta$) parameters with respect to the iterations of the sampling process is studied in Fig.~\ref{fig:Gibbs_posterior_estimation}. The parameter estimates attain convergence in fewer than $50$ iterations using Gibbs sampling in combination with Newton’s method, as described in Algorithm~\ref{algo:Gibbs Newton}. This demonstrates the efficiency of the proposed estimation scheme, yielding the converged values $(\widehat{\alpha}, \widehat{\beta}) = (0.1273,7682.7)$.

It is interesting to note that the likelihood and posterior optimization yield comparable estimates for \(\alpha\), whereas the estimates for \(\beta\) differ significantly. As proved in Appendix~C, the rate parameter (\(\beta\)) exhibits sensitivity to the signal power within the RD segment. Since signal power is governed by the RRE, the value of \(\beta\) for a given RD segment can vary with target distance and RCS. This observation reinforces the rationale for employing skewness in~\eqref{eq:skewness}---a metric dependent solely on the shape parameter---as a robust detection criterion.

\subsection{Performance analysis of skewness-based detection within the pipeline}
\begin{figure}[htbp]
    \centering
    \vspace{-0.1cm}
    \includegraphics[trim={0.2cm 0.5cm 0cm 0.25cm},width=0.485\textwidth]{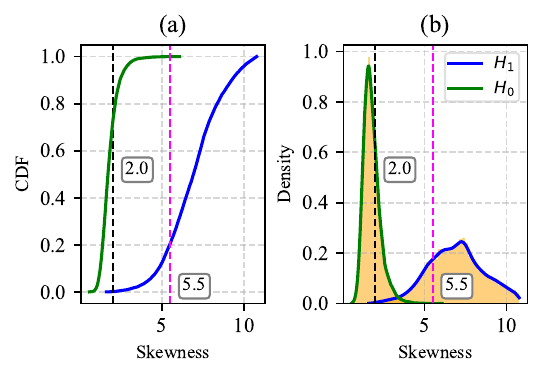}
    \caption{CDF and KDE of skewness for $H_0,H_1$ (a) CDF (b) KDE plot.}
    \label{fig:CDF_KDE_skew}
\end{figure}

The performance of the test statistic can be evaluated by studying its distribution under the two hypotheses, i.e., $p(\kappa\mid H_0)$ and $p(\kappa\mid H_1)$. With the available dataset, we observe the cumulative distribution function (CDF) and the kernel density estimator (KDE) in Fig.~\ref{fig:CDF_KDE_skew} for both hypotheses. It is first observed that $p(\kappa\mid H_0)$ is centered around a skewness value of $2$ (black dotted line), aligning with the skewness value of the exponential distribution. In contrast, $p(\kappa\mid H_1)$ consistently exhibits a large skewness value around $5.5$ (magenta dotted line), which corresponds with the estimated scale parameter value of $\widehat{\alpha}=0.13$. 

\begin{figure*}[!t]
    \centering     
    \includegraphics[ width=\textwidth]{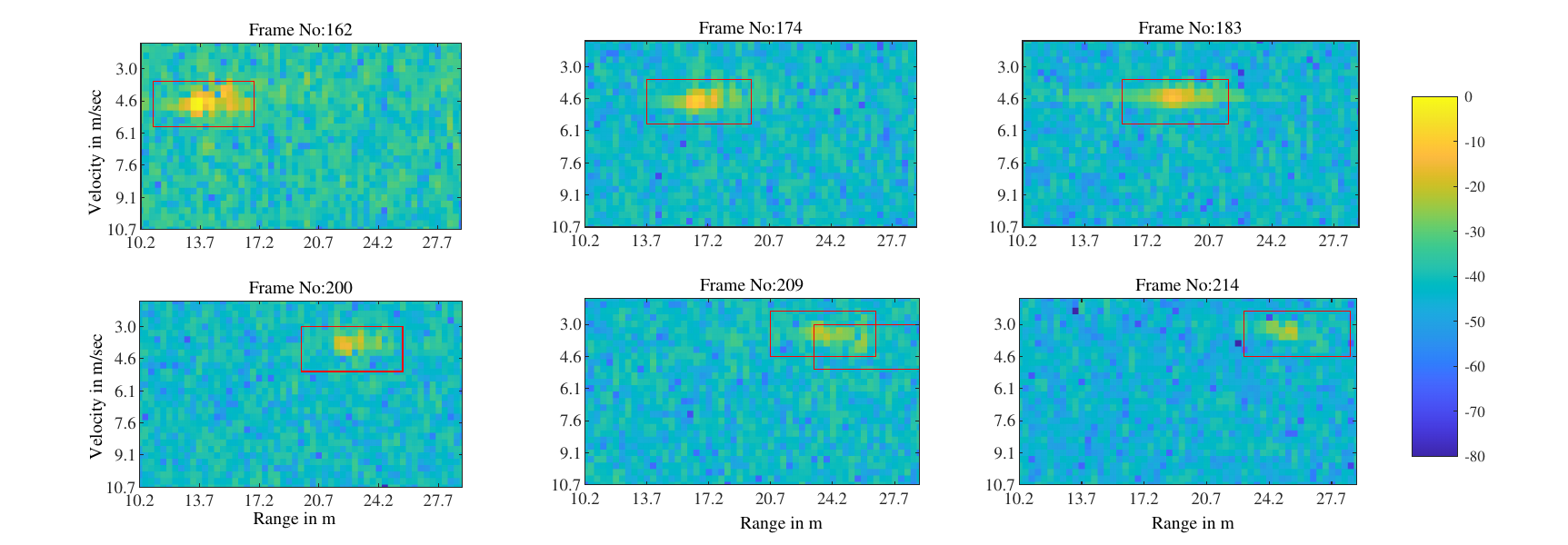}    
    \caption{Skew-based detection pipeline for different frames on Real data.(Colorbar: Power level in dB).}    
    \label{fig:skew_det_pipeline}    
\end{figure*}

Importantly, the two distributions exhibit clear separation with only marginal overlap, which facilitates the selection of a skewness-based threshold that balances the false-alarm rate ($P_{\mathrm{FA}}$) against the detection probability ($P_{\mathrm{D}}$). To determine an appropriate threshold $F(\theta) \gtrless T$, the trade-off between $P_{\mathrm{D}}$ and $P_{\mathrm{FA}}$ is evaluated within the proposed detection pipeline in Fig.~\ref{fig:skew_det_pipeline}. To assess the overall detection pipeline, an independent RD map dataset is simulated under diverse SNR conditions, distinct from those used for MLE and Gibbs sampling. Multiple RD segments, representing the same extended target with slight marginal shifts, are aggregated using IoU-based post-processing.

\begin{figure}[htbp]
    \centering    
    \includegraphics[width=0.488\textwidth,height=0.45\textwidth]{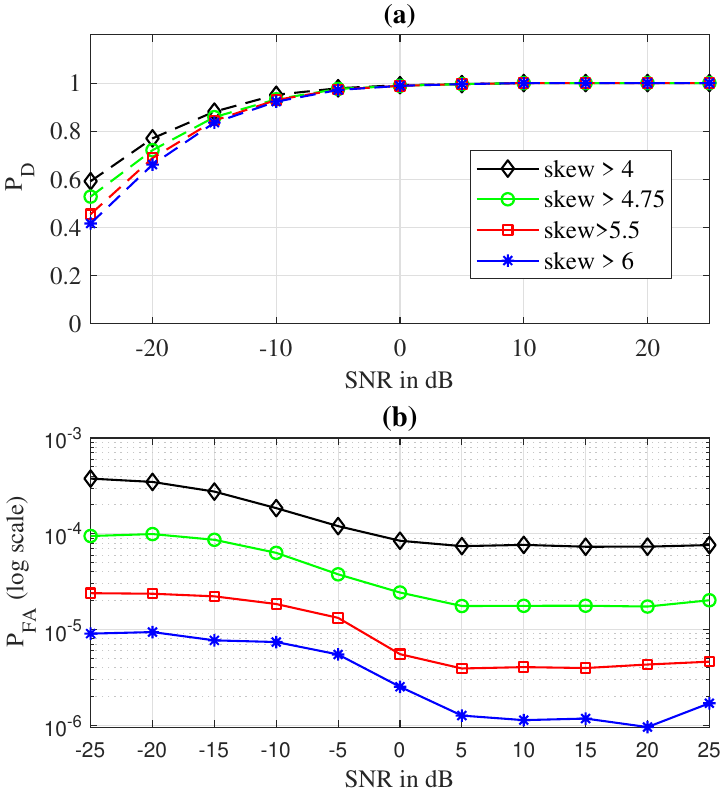}          
    \caption{Performance of skewness-based detection studied against SNR: (a) $P_{\mathrm{D}}$, (b) $P_{\mathrm{FA}}$.}
    \label{fig:skew_diff_thresh}    
\end{figure}

Fig.~\ref{fig:skew_diff_thresh} presents the probability of detection and false alarm for thresholds $T \in \{{4,4.75,5.5,6\}}$. A threshold of $T=4$ is observed in Fig.~\ref{fig:skew_diff_thresh}(a) to have superior detection probability, at the expense of an elevated false-alarm probability observed in Fig.~\ref{fig:skew_diff_thresh}(b). In comparison, $T = 4.75, 5.5, 6$ exhibit superior $P_{\mathrm{FA}}$ with marginal deterioration in $P_{\mathrm{D}}$. Across the practical SNR range of $-5$ to $20,\mathrm{dB}$, lower thresholds consistently improve $P_{\mathrm{D}}$ while increasing $P_{\mathrm{FA}}$. A threshold of $T=5.5$ provides a balanced trade-off between $P_{\mathrm{D}}$ and $P_{\mathrm{FA}}$, and is therefore chosen as the operating point for subsequent analysis.

\begin{figure}[htbp]
    \centering    
    \includegraphics[width=0.488\textwidth,height=0.45\textwidth]{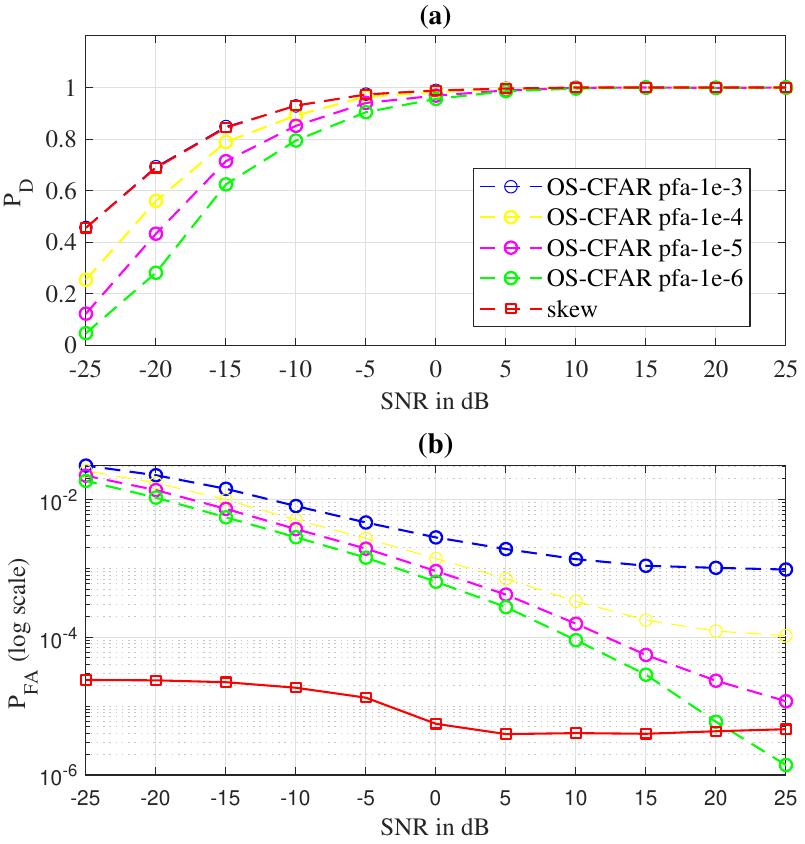}        
    \caption{Skewness-based($F(\theta) \geq 5.5$) detection vs. OS-CFAR; (a) Detection performance (b) False Alarm trend.}    
    \label{fig:Pd_Pfa_SNR}    
\end{figure}

With the chosen detection threshold for the proposed technique, it is compared with $2$D-OS-CFAR designed for $P_{\mathrm{FA}}$ varied from $10^{-3}$ to $10^{-6}$. To ensure a fair performance comparison, the CFAR window configuration is matched with the RD segment dimensions. The $P_{\mathrm{D}}$ and $P_{\mathrm{FA}}$, studied against SNR, is presented in Fig.~\ref{fig:Pd_Pfa_SNR}. The detection performance was evaluated using both simulated and real datasets. The simulated experiments comprised $350$ Monte Carlo iterations with $2$–$6$ targets placed at varying ranges and velocities, with SNR spanning $-25$ to $25$ dB.

As illustrated in Fig.~\ref{fig:Pd_Pfa_SNR}(a), the skewness-based detection exhibits superior performance, even under harsh conditions with negative SNRs. Within the pipeline, the detection performance of the proposed technique is observed in Fig.~\ref{fig:Pd_Pfa_SNR}(a) to match OS-CFAR designed for $P_{\mathrm{FA}}=10^{-3}$ and outperform OS-CFAR with $P_{\mathrm{FA}}\geq 10^{-4}$. Despite such high detection rate, as can be seen in Fig.~\ref{fig:Pd_Pfa_SNR}(b), the false alarm-rate is much lower than all the variants of OS-CFAR. It is only at SNR levels above 20 dB that the OS-CFAR's false alarm rate of $10^{-6}$ falls below that of the skewness-based detection.

The proposed technique was validated using experimental data collected with a TI radar, whose specifications are summarized in Table~\ref{tab:Realworld_spec}. Multiple acquisition trials were conducted by varying the relative orientation of the vehicle with respect to the radar. In total, $230$ dwell frames were recorded, yielding a sufficient number of RD segments for a comprehensive evaluation of the detector. Figure~\ref{fig:skew_det_pipeline} illustrates the detections produced by the proposed method (highlighted by red rectangular boxes) across a sequence of frames from one acquisition.

Detection is performed for all shifts of the RD segments across the RD map, but post-processing based on peak-centered alignment and the IoU criterion effectively suppresses redundant RD segments corresponding to the target spread. An exception is observed in Frame~$209$, where two RD segments with less than $40\%$ overlap still represent the same target.  Overall, the proposed IoU-based post-processing substantially simplifies the subsequent association stage before tracking.

\begin{figure}[htbp!]
    \includegraphics[width=0.487\textwidth]{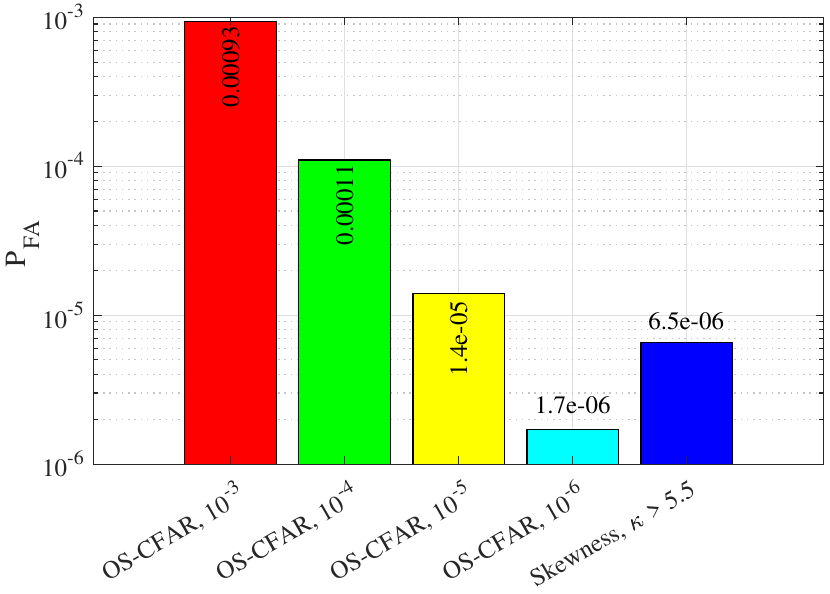}
    \caption{Skew-statistics performance against OS-CFAR for real data.}
    \label{fig:skew_real_data}
\end{figure}

In the acquired dataset, the target was detected by both OS-CFAR and the proposed technique in most frames. Accordingly, the comparative analysis emphasizes the false alarms generated by the two methods. Figure~\ref{fig:skew_real_data} shows the $P_{\mathrm{FA}}$ comparison between OS-CFAR and the proposed skewness-based detection pipeline, indicating that the latter achieves a false-alarm rate comparable to the OS-CFAR designed for $P_{\mathrm{FA}}=10^{-6}$. This performance is primarily attributed to the effectiveness of the skewness criterion in identifying suitable RD segments, complemented by the subsequent post-processing steps. Thus, the skewness-based pipeline outperforms OS-CFAR, achieving detection rates equal to or exceeding OS-CFAR at $P_{FA} = 10^{-3}$, while maintaining a low false alarm rate between $10^{-5}$ and $10^{-6}$. This performance is achieved with single dwell and without the support of association and tracking, making it highly suitable for automotive target detection.

\begin{figure}[H]
    \centering       
    \includegraphics[trim={0 0.4cm 0 0.4cm},width=0.485\textwidth]{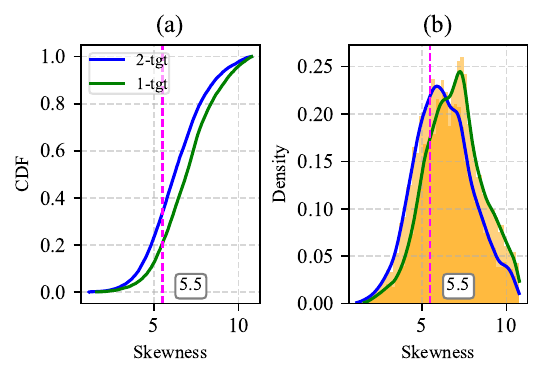}  
    \caption{CDF and KDE of skewness comparison for targets within RD-segments (a) CDF (b) KDE }    
    \label{fig:CDF and KDE of 2closely spaced targets}    
\end{figure}

In practical scenarios, the employed range-Doppler (RD) segment dimensions of \(5.98\,\mathrm{m} \times 2.13\,\mathrm{m/s}\) may encompass either a single extended target or multiple closely-spaced targets. As a result, extended or large targets can be influenced by residual scattering spread from neighboring targets, complicating the task of accurately distinguishing the number of targets present. Nevertheless, detecting the presence of target responses within the RD segment remains feasible using the proposed technique, while deferring the explicit determination of the number of targets to the tracking stage. To investigate this, the CDF and KDE of the test statistic (skewness) under hypothesis \(H_1\) are analyzed for RD segments containing two closely-situated targets, as shown in Fig.~\ref{fig:CDF and KDE of 2closely spaced targets}, and compared with those obtained for a single extended target.

It is evident that the conditional probability density \( p(\kappa \mid H_1) \) in the presence of two closely situated targets is slightly shifted to the left compared to the distribution obtained for extended targets, indicating that the same test statistic can be effectively employed for target detection. From Fig.~\ref{fig:CDF and KDE of 2closely spaced targets}(a), it can be observed that detection experiences a marginal decline in its performance for the threshold $T=5.5$. These results suggest that the proposed technique successfully identifies the presence of targets within the RD segment, while delegating the resolution of multiple closely-spaced targets to the subsequent tracking module over multiple dwells.

Thus, the pipeline effectively identifies RD segments containing response from extended targets or multiple targets by observing their skewness. Integrating this function with centering and IoU enhances detection accuracy and reduces false alarms within a single dwell. The RD segment formulation also aids in simpler association and tracking module, reducing the complexity and making pipeline highly suitable for automotive target detection.

\section{Conclusion}
This work introduces a statistically grounded RD segment-based framework for extended target detection in mmWave automotive radar, addressing limitations of conventional cell-based methods like CFAR, which often result in incomplete detections for extended targets. The proposed approach preserves the target's scattering structure by statistically characterizing RD cells using Gamma distribution modeling via MLE and Gibbs MCMC sampling. A skewness-based test statistic, robust to SNR variations, facilitates binary hypothesis classification. The detection pipeline incorporates peak-centered segment alignment and IoU to handle multiple RD segments within a single dwell, simplifying track formation by reducing association complexity. Simulations and experimental results demonstrate reliable detection performance with minimal false alarm rates compared to OS-CFAR.
\begin{appendices}
{\small
\section{Expressions for NLL and their gradients}
Let $z_1,z_2,....z_\mathbb{N}$ be i.i.d. and drawn from a Gamma distribution with unknown $(\alpha,\beta), z_j \geq 0$,
\begin{equation}
    \begin{split}
        f_z(z_{1}, z_{2}, \dots, z_{\mathbb{N}} \mid H_1) &\simeq \prod_{j=1}^{\mathbb{N}} \frac{\beta ^{\alpha}}{\Gamma(\alpha)} z_{j}^{\alpha - 1} \exp{(-\beta z_{j})}       
    \end{split}
     \label{eq:gamma_unimodal_posterior}
\end{equation}

\begin{equation}
    \begin{split}
    f_z(z_{1}, z_{2}, \dots, z_{\mathbb{N}} \mid H_1) &\simeq \frac{\beta ^{\mathbb{N}\alpha}}{(\Gamma(\alpha))^\mathbb{N}} \exp{\bigg(-\beta \sum_{j=1}^\mathbb{N}z_{j}\bigg)} \prod_{j=1}^{\mathbb{N}}  z_{j}^{\alpha - 1} .
    \label{eq:Ngamma_variables}
    \end{split}
\end{equation}

The corresponding Log-likelihood function is given by
\begin{equation}
    \begin{split}
     \mathcal{L}(z;\alpha,\beta) = \log(f_z(z_{1}, z_{2}, \dots, z_{\mathbb{N}} \mid H_1))\\
     =\mathbb{N}\alpha \log(\beta) - \mathbb{N}\log(\Gamma(\alpha)) + (\alpha -1) \sum_{j=1}^\mathbb{N} \log(z_j) -\beta\sum_{j=1}^\mathbb{N}z_j .
    \label{eq:log_Ngamma}    
    \end{split}
\end{equation}
The gradients w.r.t. the parameters are obtained as
\begin{align}
    \nonumber \frac{\partial}{\partial\alpha}\mathcal{L}(z;\alpha,\beta)&= \sum_{j=1}^{\mathbb{N}} (\log(\beta) - \frac{\Gamma'(\alpha)}{\Gamma(\alpha)} + \log(z_j)) \\ &= \mathbb{N}(\log(\beta) - \Psi(\alpha)) + \sum_{j=1}^{\mathbb{N}} \log(z_j)
     \label{eq:grad_alpha}
\end{align}

\begin{align}
    \frac{\partial}{\partial\beta}\mathcal{L}(z;\alpha,\beta) &= \frac{\mathbb{N}\alpha}{\beta} - \sum_{j=1}^\mathbb{N} z_j
     \label{eq:grad_alpha_beta}
\end{align}

The MLE estimates of the parameters are arrived as ~\cite{Papoulis2002probability}, 
\begin{align}
    \widehat{\beta}_{\mathrm{ML}} &= \frac{\widehat{\alpha}_{\mathrm{ML}}}{\bar{z}} 
    \label{eq:MLE_beta}\\
    \log(\widehat{\alpha}_{\mathrm{ML}}) - \Psi(\alpha) &= \log(\bar{z}) - \frac{\sum_{j=1}^\mathbb{N}\log(z_j)}{\mathbb{N}} , 
    \label{eq:MLE_beta_alpha}
\end{align}

where $\Psi(\alpha) =\frac{\Gamma'(\alpha)}{\Gamma(\alpha)}$ is the digamma function, $\bar{z} = \frac{\sum_{j=1}^\mathbb{N} z_j}{\mathbb{N}}$ is the sample mean.

\begin{align}
         \frac{\partial^2}{\partial\alpha^2} \mathcal{L}(z;\alpha,\beta) &= -\mathbb{N}\Psi'(\alpha).
         \label{eq:Hess_alpha}
\end{align}

Here $\Psi'(\alpha) = \frac{d(\Psi(\alpha))}{d\alpha}$, polygamma function of the order one.
\section{Gamma Posterior derivation for the rate parameter}
According to Bayes’ theorem, the posterior distribution is proportional to the product of the likelihood and the prior:
{\small
\begin{align}
    Posterior \propto Likelihood \times Prior
\end{align}
}
From the list of conjugate priors, we know that the Gamma distribution is conjugate to itself for the rate parameter $\tilde{beta}$ under the assumption of known shape parameter $\tilde{\alpha}$. That is, if the rate parameter $\tilde{\beta}$ is assigned a Gamma prior distribution, and the likelihood also follows a Gamma distribution parameterized by $\tilde{\beta}$, then the resulting posterior distribution for $\tilde{\beta}$, is a Gamma distribution~\cite{lambertstudent,jeffreys1998theory,fink1997compendium,wikipedia:conjugateprior}.
Consider the prior distribution for the rate parameter $\tilde{\beta}$, with their corresponding Gamma parameters $(a,b)$:
\begin{align}
     \tilde{\beta} \sim \text{Gamma}(a,b) =f_{\tilde{\beta}}(a,b)&= \frac{b^a {\tilde{\beta}}^{(a-1)} e^{-b{\tilde{\beta}}}}{\Gamma(a)},
     \label{eq:gamma_prior}
\end{align}
and with known shape parameter $\tilde{\alpha}$, the Likelihood for the rate parameter $\tilde{\beta}$ leveraging from~\eqref{eq:Ngamma_variables},
\begin{align}
    f_z((z,\tilde{\alpha}) \mid \tilde{\beta}) &= \frac{{\tilde{\beta}} ^{\mathbb{N}{\tilde{\alpha}}}}{(\Gamma({\tilde{\alpha}}))^\mathbb{N}} \exp{(-{\tilde{\beta}} \sum_{j=1}^\mathbb{N}z_{j})} \prod_{j=1}^{\mathbb{N}}  z_{j}^{{\tilde{\alpha}} - 1}.
    \label{eq:gamma_likelhood}
\end{align}
The posterior for the rate parameter $\tilde{\beta}$ can be derived by multiplying~\eqref{eq:gamma_prior} and ~\eqref{eq:gamma_likelhood},
\begin{align}
    \nonumber f_z({\tilde{\beta}}\mid (z,\tilde{\alpha})) &\propto  f_z(z \mid {\tilde{\alpha}},{\tilde{\beta}}) f_{\tilde{\beta}}(a,b)\\
    \nonumber f_z({\tilde{\beta}}\mid (z,\tilde{\alpha})) &\propto \frac{{\tilde{\beta}} ^{\mathbb{N}{\tilde{\alpha}} + a -1}}{(\Gamma({\tilde{\alpha}}))^\mathbb{N}\Gamma(a)} e^{-{\tilde{\beta}} (\sum_{j=1}^\mathbb{N}z_{j}+b)} \prod_{j=1}^{\mathbb{N}}  z_{j}^{{\tilde{\alpha}} - 1} \\
    \nonumber f_z({\tilde{\beta}}\mid (z,\tilde{\alpha})) &\propto \frac{{\tilde{\beta}} ^{\mathbb{N}{\tilde{\alpha}} + a -1}}{(\Gamma({\tilde{\alpha}}))^\mathbb{N}\Gamma(a)} e^{-{\tilde{\beta}} (\mathbb{N}\bar{z}+b)} \prod_{j=1}^{\mathbb{N}}  z_{j}^{{\tilde{\alpha}} - 1} \\
    f_z({\tilde{\beta}} \mid (z,\tilde{\alpha})) &\propto \mathrm{Gamma}(\mathbb{N}{\tilde{\alpha}} + a,\mathbb{N}\bar{z}+b)
    \label{eq:posterior_for_gibbs}
\end{align}
}
\section{Normalization effect on Gamma distribution parameters}
When the measurements $z_1, z_2, \ldots, z_{\mathbb{N}}$ are scaled by a factor $\lambda$ that depends on the target distance, RCS, and the SNR, the mean is correspondingly scaled as $E[z/\lambda] = \overline{z}/\lambda$. The associated Gamma distribution parameters can then be determined from~\eqref{eq:MLE_beta_alpha} and~\eqref{eq:MLE_beta} as,
\begin{align}
    \log(\tilde{\alpha}) - \Psi(\tilde{\alpha}) &= \log(\bar{z}/\lambda) - \frac{\sum_{j=1}^\mathbb{N}\log(z_j/\lambda)}{\mathbb{N}} \nonumber \\
     \log(\tilde{\alpha}) - \Psi(\tilde{\alpha}) &= \log(\bar{z}) - log(\lambda) - \frac{\sum_{j=1}^\mathbb{N}\log(z_j) - \mathbb{N}log(\lambda)}{\mathbb{N}} \nonumber\\
     \log(\tilde{\alpha}) - \Psi(\tilde{\alpha}) &= \log(\bar{z}) - \frac{\sum_{j=1}^\mathbb{N}\log(z_j)}{\mathbb{N}}\\
     \tilde{\alpha} &= \alpha \quad \text{(invariant under scaling)}
     \label{eq:normalization_alpha}
\end{align}
and
\begin{align}
        \tilde{\beta} &= \frac{\tilde{\alpha}}{E[z/\lambda]} = \frac{\lambda\tilde{\alpha}}{\bar{z}} = \lambda \beta. \quad \text{(varies under scaling)}                 
        \label{eq:normalization_beta}
\end{align}
\end{appendices}
\begingroup\footnotesize 
\bibliographystyle{IEEEtran} 
\bibliography{references.bib}
\endgroup
\newpage
\end{document}